\documentclass[reprint,aps,pra,floatfix]{revtex4-2}
\usepackage{amssymb,amsmath,amsfonts}
\usepackage{graphicx}
\usepackage{appendix}
\usepackage[dvipsnames]{xcolor}
\usepackage{bm,bbm}
\usepackage{dcolumn}
\usepackage{mathrsfs}
\usepackage{calc}
\usepackage{color}

\usepackage[normalem]{ulem}
\definecolor{nir_pink}{RGB}{212, 55, 172}

\usepackage[colorlinks=true, citecolor=teal, urlcolor=purple,linkcolor=purple]{hyperref}

\begin{document}

\title{Stochastic Resetting vs. Thermal Equilibration: Faster Relaxation, Different Destination}
% \title{Stochastic Resetting Relaxes Faster than Thermal Equilibration but to a Different Steady-state}

\author{Nir Sherf}
\affiliation{Raymond \& Beverly Sackler School of Chemistry, Tel Aviv University, Tel Aviv 6997801, Israel}

\author{R\'emi Goerlich}
\affiliation{Raymond \& Beverly Sackler School of Chemistry, Tel Aviv University, Tel Aviv 6997801, Israel}

\author{Barak Hirshberg}
\email{hirshb@tauex.tau.ac.il}
\affiliation{Raymond \& Beverly Sackler School of Chemistry, Tel Aviv University, Tel Aviv 6997801, Israel}
\affiliation{Center for Computational Molecular and Materials Science, Tel Aviv University, Tel Aviv 6997801, Israel.}
\affiliation{The Center for Physics and Chemistry of Living Systems, Tel Aviv University, Tel Aviv 6997801, Israel.}

\author{Yael Roichman}
\email{roichman@tauex.tau.ac.il}
\affiliation{Raymond \& Beverly Sackler School of Physics and Astronomy, Tel Aviv University, Tel Aviv 6997801, Israel}
\affiliation{Raymond \& Beverly Sackler School of Chemistry, Tel Aviv University, Tel Aviv 6997801, Israel}
\affiliation{The Center for Physics and Chemistry of Living Systems, Tel Aviv University, Tel Aviv 6997801, Israel.}

\date{\today}
\begin{abstract}
Stochastic resetting is known for its ability to accelerate search processes and induce non-equilibrium steady states.
Here, we compare the relaxation times and resulting steady states of resetting and thermal relaxation for Brownian motion in a harmonic potential.
We show that resetting always converges faster than thermal equilibration, but to a different steady-state. The acceleration and the shape of the steady-state are governed by a single dimensionless parameter that depends on the resetting rate, the viscosity, and the stiffness of the potential. 
We observe a trade-off between relaxation speed and the extent of spatial exploration as a function of this dimensionless parameter. 
Moreover, resetting relaxes faster even when resetting to positions arbitrarily far from the potential minimum.
\end{abstract}

\maketitle

\textit{Introduction} --- 
%- SR is well-known to lead to acceleration in first-passage processes. It is also known to induce NESS, as long as the mean time between resetting events is finite. 
Stochastic resetting stops a random process and re-initializes it with independent initial conditions \cite{evans_diffusion_2011,evans_stochastic_2020, manrubia_stochastic_1999}.
It is well known to accelerate search processes \cite{eliazar_searching_2007, evans_diffusion_2011, evans_diffusion_2011-1, evans_diffusion_2014, kusmierz_first_2014, kusmierz_optimal_2015, bhat_stochastic_2016, pal_first_2017, pal_first_2019, bressloff_search_2020, singh_backbone_2021, tal-friedman_experimental_2020, pal_random_2024}, 
to enable optimization of queuing strategies \cite{bressloff_queueing_2020, pal_inspection_2022, bonomo_mitigating_2022,roy_queues_2024}, 
to enhance sampling in molecular simulations \cite{blumer_combining_2024, blumer_stochastic_2022, blumer_have_2025}, 
and even to control the yield of chemical reactions \cite{reuveni_role_2014, scher_unified_2021, pal_inspection_2022, bressloff_diffusion-mediated_2022, biswas_rate_2023}.
Stochastic resetting can be applied to various types of dynamics under diverse conditions, including in the presence of external potentials \cite{pal_diffusion_2015, evans_run_2018, gupta_stochastic_2018}.
In addition, resetting is known to induce a non-equilibrium steady state (NESS) if the mean time between resetting events is finite \cite{evans_diffusion_2011, gupta_fluctuating_2014, durang_statistical_2014, singh_general_2022}.
Several works have focused on characterizing the resulting NESS, whose probability density distribution significantly differs from the Boltzmann distribution in thermal equilibrium and depends on the resetting rate, the resetting protocol, and the underlying potential \cite{evans_diffusion_2011, gupta_fluctuating_2014,evans_diffusion_2014, pal_diffusion_2015, pal_diffusion_2016, evans_run_2018}.

Much less is known about the \textit{rate of relaxation} towards the NESS, especially in comparison to thermal relaxation. 
For example, the full time-dependent probability distribution is analytically known only for free diffusion under resetting \cite{pal_diffusion_2015}. 
Since solving the dynamics is often not analytically tractable, several approximations were previously used to describe how the NESS emerges \cite{majumdar_dynamical_2015, gupta_stochastic_2018, singh_resetting_2020, singh_backbone_2021, singh_general_2022, garcia-valladares_stochastic_2024}. 
Usually, the stationary distribution is established near the resetting point first and then spreads out with time. Even for a diffusing particle trapped in a harmonic potential the evolution of the steady state and the rate of relaxation remains largely unexplored \cite{pal_diffusion_2015,trajanovski_ornstein-uhlenbeck_2023}.

It was recently demonstrated that stochastic resetting can accelerate a transition between two equilibrium states in a V-shaped potential \cite{goerlich_resetting_2024}. 
This was achieved by turning off the first potential, initiating stochastic resetting for a finite time, and turning back on the second potential. 
It was possible because the steady state distribution for stochastic resetting of a freely diffusing particle has the same form as the Boltzmann equilibrium distribution in a V-shaped potential. 
By tuning the resetting rate, the authors obtained the same target equilibrium state, only faster. 
This result raises an interesting question: does stochastic resetting always relax to its steady state faster than thermal equilibration? If so, stochastic resetting could accelerate state-to-state transformations in heat engines \cite{martinez2017colloidal, lahiri2024efficiency}, or bypass thermalization in mechanical relaxations \cite{martinez2016engineered, le2016fast}.

Here, we address this question by comparing the relaxation to steady state of a diffusing particle in a harmonic potential with and without stochastic resetting. 
We find that the system under resetting always relaxes faster than its thermally equilibrated counterpart, but to a different NESS. 
Remarkably, this speedup is governed by a single dimensionless parameter, which depends on the stiffness of the potential, the viscosity, and the resetting rate, effectively characterizing the dynamical behavior of the system. 
We also find a tradeoff between the speedup and the extent of spatial exploration as a function of this dimensionless parameter. 
Finally, we show that the approach to steady state is dictated by the slowest relaxing moment. For resetting near the potential minimum, the second moment dominates the dynamics while for resetting far from the minimum, the first moment becomes dominant. Regardless of the resetting point, under stochastic resetting the system always relaxes faster than thermal equilibration.\\

\textit{Model} --- 
We study the relaxation towards steady state for a Brownian particle confined in a one-dimensional harmonic potential under stochastic resetting (SR) and compare it to thermal relaxation. 
The dynamics of the trapped particle obeys the Ornstein–Uhlenbeck (O-U) process
\begin{equation}
    \dot{x} = -\omega_0 x + \sqrt{2D}~ \eta(t)
    \label{eq:langevin}
\end{equation}
where $\omega_0 = \kappa/\gamma$ is the inverse characteristic timescale of the process, $\kappa$ is the stiffness of the harmonic potential and $\gamma$ is the Stokes drag coefficient. $D = k_{\rm B} T / \gamma $ is the diffusion coefficient,  $k_{\rm B}$ is the Boltzmann constant, and $T$ is the bath temperature. 
Here, $\eta(t)$ represents thermal white-noise with zero mean.
Stochastic resetting is then applied by resetting the particle's position at random time intervals taken from a Poisson distribution with a constant mean rate $r$. 
At each resetting event the particle is instantaneously returned to a predetermined position $x_0$.

\begin{figure}[htbp]
    \centering
    \includegraphics[width=1\linewidth]{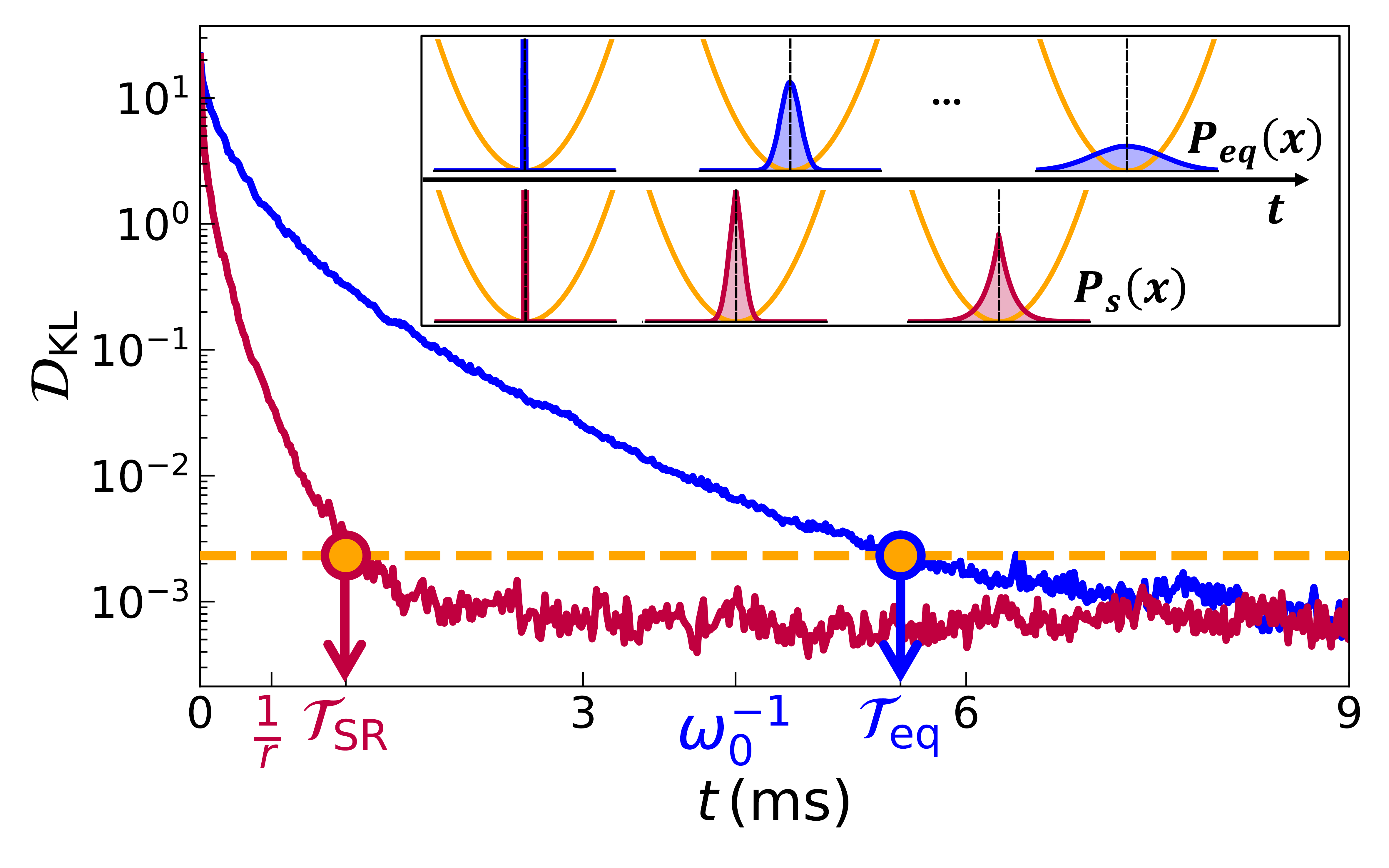}
    \caption{KLD as a function of time for thermal equilibration (blue) and SR (red, $r=1.8 \,\mathrm{kHz} = 7.5 \omega_0$) using trap stiffness $\kappa = 2 \, \mathrm{pN /\mu m}$. 
    The relaxation times (yellow circles) are determined when the KLD first drops below a threshold (dashed yellow line). 
    This allows us to measure $\mathcal{T}_{\rm eq} =5.49 \, \mathrm{ms} = 1.31\omega_0^{-1}$ and $\mathcal{T}_{\rm SR} =1.14 \, \mathrm{ms}= 0.27\omega_0^{-1}$.
    The inset graphically shows the evolution of the probability distribution for both cases.}
    \label{fig:relaxation time}
\end{figure}

Without resetting, the relaxation towards equilibrium is well known \cite{risken1989fokker}.
Specifically, the time-dependent probability distribution $P(x,t)$ is described by a Gaussian propagator with variance $\sigma^2(t)$ and mean $\mu(t)$, 
\begin{equation}
  P(x,t)=\sqrt{\frac{1}{2\pi\sigma^2(t)}}e^{-(x-\mu(t))^2/2\sigma^2(t)}.
  \label{eq:propagator}
\end{equation}
The temporal evolution of a probability distribution starting at $P(x, 0)=\delta (x-x_0)$, is fully determined by the relaxation of its first two moments, the variance $\sigma^2(t)= k_{\rm B} T/\kappa~[1-\exp(-2 \omega_0 t)] $ and the mean $\mu(t) = x_0[1-\exp(-\omega_0 t)]$. At long times, we obtain the Boltzmann distribution of a particle in a harmonic trap.
In contrast, the full propagator for an O-U process under resetting is unknown. However, it reaches a steady-state $P_{\rm s}(x)$ whose analytical form is given in the SI, section~\ref{app:SR steady-state}, and differs from the Boltzmann equilibrium distribution (see the inset of Fig.~\ref{fig:relaxation time}) \cite{pal_diffusion_2015, trajanovski_ornstein-uhlenbeck_2023, goerlich_taming_2024}.
%Furthermore, $P(x,t)$ is generally not Gaussian anymore, and unlike the thermal relaxation, its evolution is not captured by the first two moments $\mu(t)$ and $\sigma^2(t)$.

%To study the relaxation to steady state under resetting, we numerically solve Eq.~(\ref{eq:langevin}) (see Appendix \ref{app:simulations} for details).
To study the relaxation to steady state under resetting, we perform overdamped Langevin simulations~\cite{volpe_simulation_2013,pal_diffusion_2015} and subject them to SR. 
Resetting times are drawn from an exponential distribution with rate parameter $r$. 
The simulations are done in arbitrary units where $dt=0.1, \gamma=500,kT=1$.
The parameters $\gamma, \kappa$ and $T$ are chosen to be consistent with a colloidal particle of radius $0.5 \,\mu \mathrm{m}$ in water at $T=300\, \mathrm{K}$ trapped in optical tweezers. 
This means that the timestep is equivalent to $1.677 \,\mu \mathrm{s}$ and potential of $\kappa=1$ is equivalent to $1\, \mathrm{pN/}\mu \mathrm{m}$, typical values that indicate we are indeed at the overdamped regime.
We run at least $10^5$ trajectories of $10^4$ steps in all cases.

We start our analysis by defining the relaxation time to steady-state $\mathcal{T}$ through the Kullback-Leibler Divergence (KLD), which measures the time-dependent distance between the evolving state $P(x,t)$ and the final steady-state $P_{\rm s}(x)$ as $\mathcal{D}_{\rm KL}(t) = \int P(x,t) \ln[P(x,t) / P_{\rm s}(x)] \mathrm{d} x$ ~\cite{goerlich_resetting_2024}.
Specifically, $\mathcal{T}$ is the time at which the KLD falls below a chosen threshold (see Fig.~\ref{fig:relaxation time} and SI, section~\ref{app:threshold} for details).
Both $P(x,t)$ and $P_{\rm s}(x)$ are evaluated numerically as histograms from an ensemble of trajectories.
Note that when $r=0$, $P_{\rm s}(x) = P_{\rm eq}(x)$.
This definition allows us to compare on equal footing the relaxation times with resetting ($\mathcal{T}_{\rm SR}$) and without ($\mathcal{T}_{\rm eq}$), which both differ from the characteristic timescales $\omega_0^{-1}$ and $r^{-1}$.\\

\textit{Resetting accelerates relaxation} ---
Fig.~\ref{fig:relaxation time} shows the $\mathcal{D}_{\rm KL}(t)$ without resetting (blue line) and with resetting at a rate $r=1.8 \, \mathrm{kHz} = 7.5 \omega_0$ (red line).
Here, we define the \textit{speedup} induced by resetting as the ratio $\mathcal{T}_{\rm eq}/\mathcal{T}_{\rm SR}$.
For this choice of parameters, we find that resetting significantly accelerates the arrival to steady-state.
As seen in the inset, the final steady-state is also modified by resetting.
In the following, we test the effect of the physical parameters of the system (resetting rate, trap stiffness and resetting position) on the speedup.

\begin{figure}[htbp]
    \centering
    \includegraphics[width=1\linewidth]{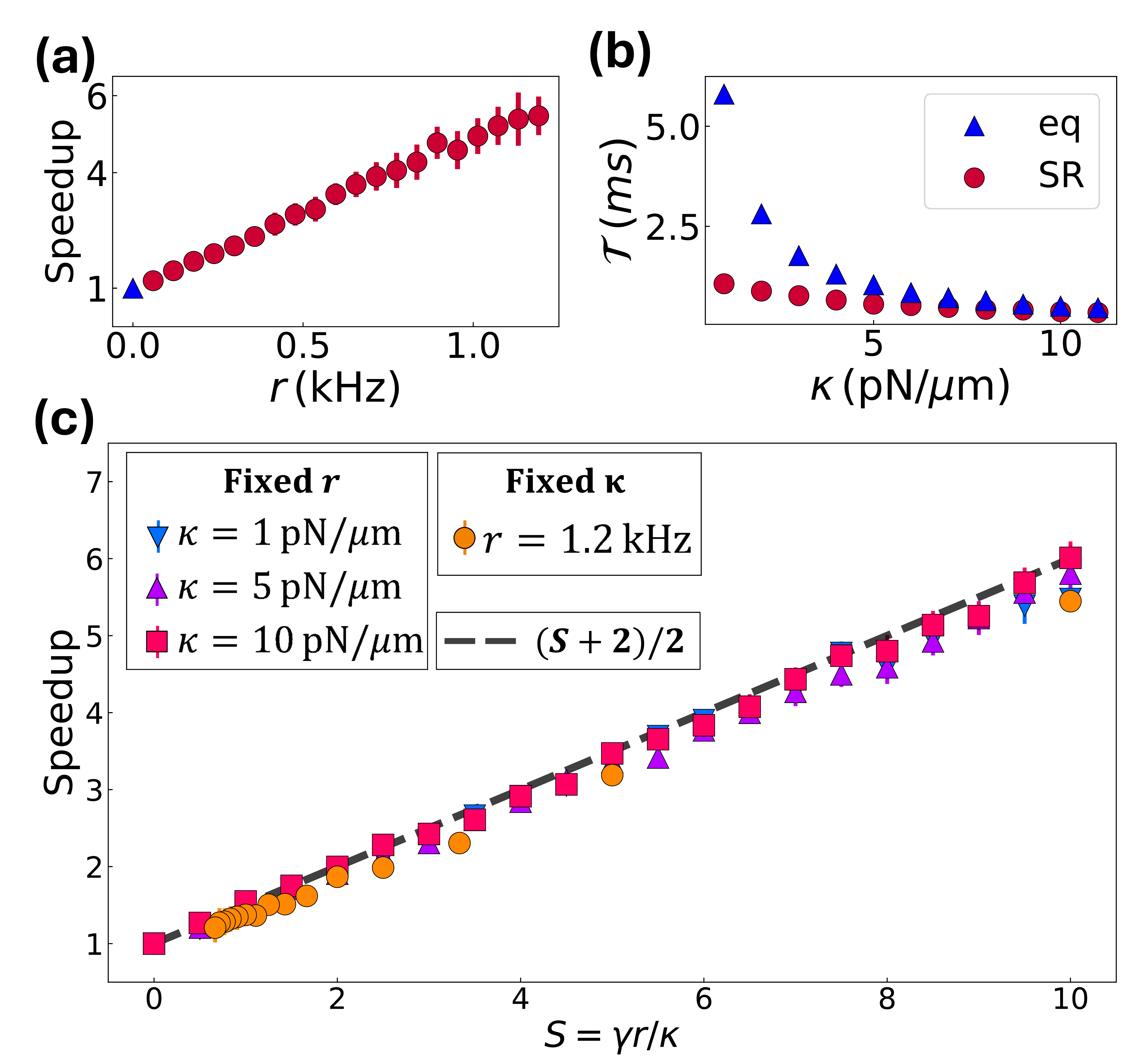}
    \caption{(a) The speedup induced by SR, $\mathcal{T}_{\rm eq}/\mathcal{T}_{\rm SR}$, against $r$ for $\kappa= 1 \, \mathrm{pN/\mu m}$. Red circles for $r\neq0$, blue triangle underlines the thermal case $r=0$ as shown in the legend of (b).
    (b) Relaxation times against $\kappa$ for $r=1.12 \, \mathrm{kHz}$. The duration of relaxation under SR is smaller than the corresponding thermal process for all $\kappa$. This effect diminishes at large $\kappa$. 
    (c) Speedup $\mathcal{T}_{\rm eq}/\mathcal{T}_{\rm SR}$, measured for various systems with different values of $r$ and $\kappa$, collapses onto a single master curve when plotted against the dimensionless parameter $S$, demonstrating that relaxation dynamics are governed solely by $S$. 
    The dashed black line is the ratio between the timescales of the second moments, showing that the speedup in relaxation of the full distribution can be characterized by the relaxation of the moments.}
    \label{fig:relaxation analysis}
\end{figure}

Fig.~\ref{fig:relaxation analysis}~(a) shows the speedup as a function of $r$ for a fixed $\kappa$.
We find that SR always accelerates relaxation since the speedup increases monotonically with the resetting rate.
Using parameters typical for an experiment of resetting in an optical trap~\cite{Besga2020, tal-friedman_experimental_2020, goerlich_taming_2024}, we see a speedup of up to $\sim 5$.
In Fig.~\ref{fig:relaxation analysis}~(b), we plot the speedup as a function of the potential stiffness $\kappa$ for a given $r$. 
We find that acceleration persists for all $\kappa$.
However, the speedup diminishes with increasing $\kappa$ as the difference between $\mathcal{T}_{\rm eq}$ and $\mathcal{T}_{\rm SR}$ decreases and the external potential dynamics dominate.

To gain further insight on the behavior of the speedup, we combine the main dynamical parameters into a single dimensionless number $S = \gamma r/\kappa$. %, which captures the two competing timescales, $r^{-1}$ and $\omega_0^{-1}$.
This parameter effectively measures the interplay between the resetting and diffusive processes on the system's evolution and NESS.
For small $S$ values, the evolution is dominated by the harmonic potential, for large $S$ it is dominated by the resetting.
As such, it is equivalent to a Péclet number in the context of resetting, a ratio of an \textit{advection}-like timescale $r^{-1}$ and a \textit{diffusion} timescale $\gamma / \kappa =\omega_0^{-1}$.
In Fig.~\ref{fig:relaxation analysis}~(c), we plot the measured speedup as a function of $S$.
Remarkably, we see that values of the speedup, measured from simulations with varying values of $r$ and $\kappa$, all collapse onto a single master curve.
This universal scaling demonstrates that the speedup in relaxation is fully determined by $S$.

To elucidate this universality, we show that the Fokker-Planck equation for a Brownian particle in a harmonic potential under resetting with rate $r$ can be recast into a dimensionless form, governed by $S$
\begin{align}
\begin{split}
    \partial_{\tilde{t}} P(\tilde{x}, \tilde{t}) &= \partial_{\tilde{x}}(\tilde{x} P(\tilde{x}, \tilde{t})) \\
    &- S\left[ (1 - \partial_{\tilde{x}}^2)P(\tilde{x}, \tilde{t}) - \delta(\tilde{x_0}-\tilde x)\right],
    \label{eq: FP(S)}
\end{split}
\end{align}
with dimensionless parameters $\tilde x = x / \alpha$ where $\alpha = \sqrt{D/r}$ is the characteristic length-scale of the free diffusing particle under resetting~\cite{evans_diffusion_2011} and $\tilde t = \omega_0 t$. 
Eq.(\ref{eq: FP(S)}) shows that indeed the evolution and NESS are both determined by $S$ alone.

We now turn to study the speedup dependence on $S$ analytically. 
Solving Eq.~(\ref{eq: FP(S)}) is generally not trivial, but we obtain further understanding of the system's relaxation dynamics by deriving an analytical expression for the time-dependent second moment,
\begin{equation}
        \langle \tilde x^2 \rangle^{SR}(\tilde t) = \frac{2S}{S+2}\left(1-e^{-(S+2)\tilde t}\right).
        \label{eq: var_sr(t),x0=0}
\end{equation}
Eq.~(\ref{eq: var_sr(t),x0=0}) shows that the relaxation of the second moment is characterized by the timescale $(S+2)^{-1}$.
In the same dimensionless form, the relaxation timescale of the second moment in the absence of resetting is $1/2$.
Fig.~\ref{fig:relaxation analysis}~(c) shows that the numerically measured speedup exactly follows the ratio of these two timescales $(S+2)/2$.
Namely, the speedup in relaxation time of the entire distribution can be described via the ratio of the two second moments alone.

\begin{figure}[htbp]
    \centering
    \includegraphics[width=1\linewidth]{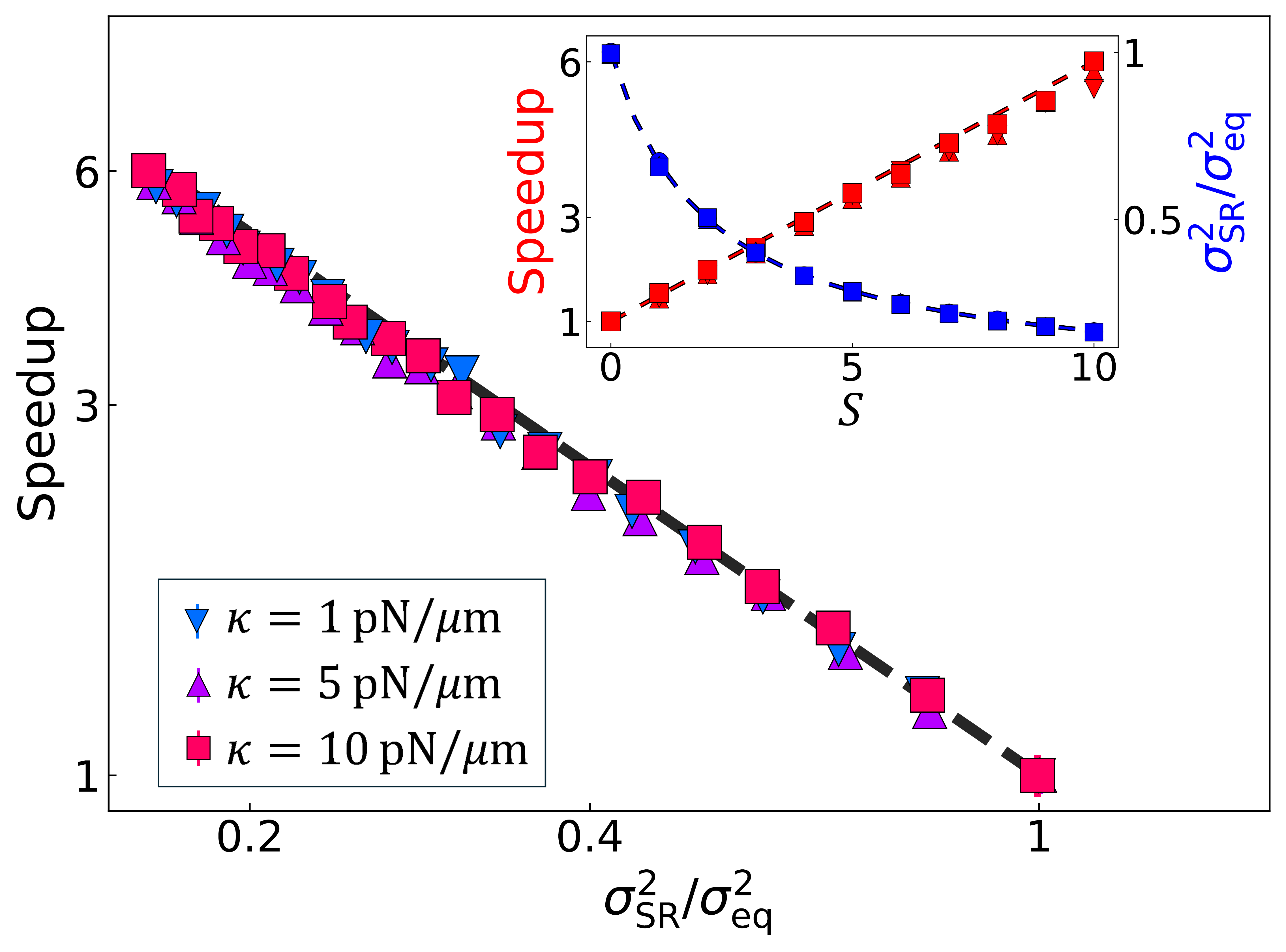}
    \caption{Speedup vs. spatial exploration shows a direct inverse relation, black dashed line follows the analytical expression $y=x^{-1}$, here $Speedup=(\sigma_{SR}^2/\sigma_{eq}^2)^{-1}$. Inset, red - speedup $\mathcal{T}_{\rm eq}/\mathcal{T}_{\rm SR}$, blue - relative spatial exploration $\sigma_{\rm SR}^2/\sigma_{\rm eq}^2$, dashed lines represent the corresponding analytical expressions red - $(S+2)/2$, blue - $2/(S+2)$.
    }
    \label{fig:tradeoff}
\end{figure}

\textit{Spatial exploration} --- As can be seen from Eq.~\ref{eq: FP(S)} and in the inset of Fig.~\ref{fig:relaxation time}, $S$ determines not only the relaxation but also the resulting NESS.
Here, we examine how the increase in speedup with $S$ affects the broadness of the NESS, as characterized by the second moment of the position. 
We define the relative \textit{spatial exploration} through the ratio between the steady-state second moments, with and without resetting  $\sigma_{\rm SR}^2/\sigma_{\rm eq}^2$ (where $\sigma^2\equiv\langle x^2(t\rightarrow \infty)\rangle $).
%To do so, we measure the extent of \textit{spatial exploration} through the ratio between the steady-state second moments, with and without resetting  $\sigma_{\rm SR}^2/\sigma_{\rm eq}^2$ (where $\sigma^2\equiv\langle x^2(t\rightarrow \infty)\rangle $).
Eq.~(\ref{eq: var_sr(t),x0=0}) implies that the ratio of spatial exploration takes the form $\sigma^2_{\rm SR}/\sigma^2_{\rm eq} = 2/(S+2)$, which is the inverse of the ratio of durations. 
Namely, the same scaling relation holds for the thermal and driven systems $\mathcal{T}_{\rm eq}/\sigma^2_{\rm eq} = \mathcal{T}_{\rm SR}/\sigma^2_{\rm SR} = 1/2S$. 
Thus, the narrower the steady-state, the faster it is reached (see Fig.~\ref{fig:tradeoff} and its inset).

\begin{figure}[htbp]
    \centering
    \includegraphics[width=1\linewidth]{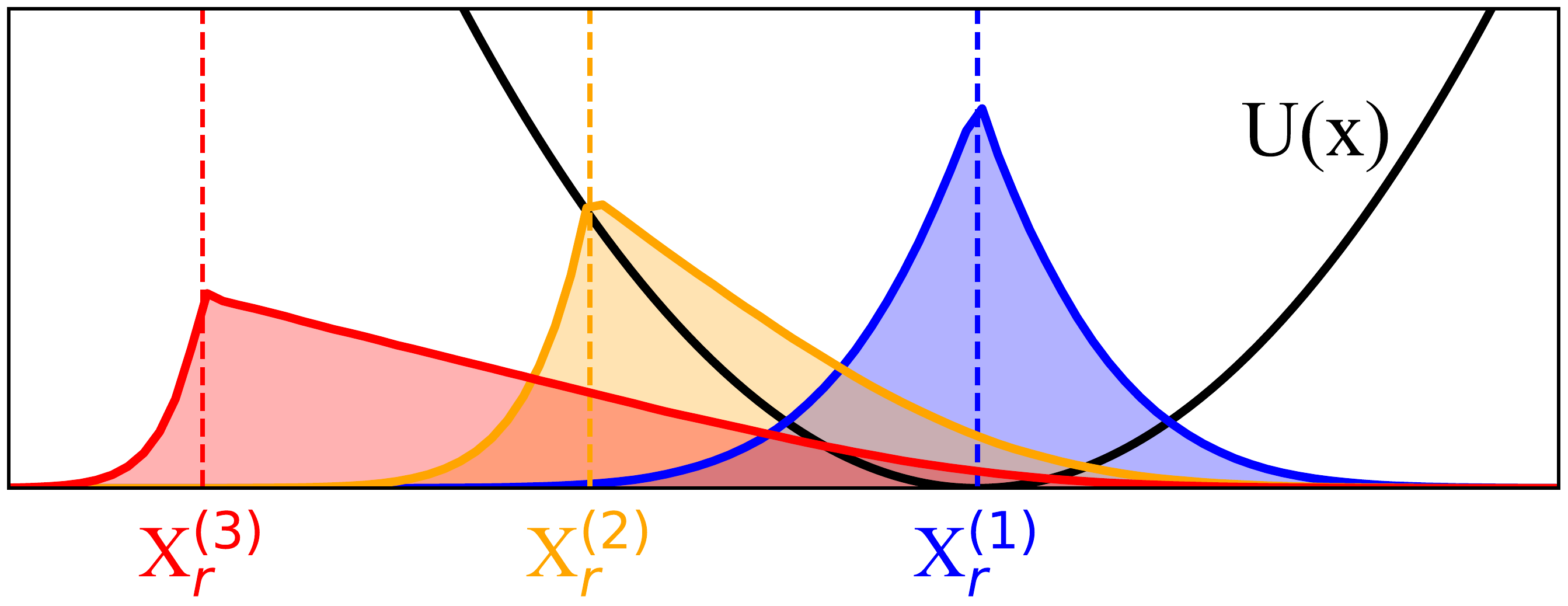}
    \caption{The NESS $P_{\rm s}(x)$of an SR process for various resetting positions from right to left: centered process $X_r^{(1)}=0$ (blue), $X_r^{(2)}\approx 2.25 \sigma$ (yellow), $X_r^{(3)}\approx 4.5\sigma$ (red). Increasing the distance between the resetting position and the center of the potential leads to skewed distributions, but also to larger spatial extension.}
    \label{fig:Xr-NESS}
\end{figure}

\begin{figure}[htbp]
    \centering
    \includegraphics[width=1\linewidth]{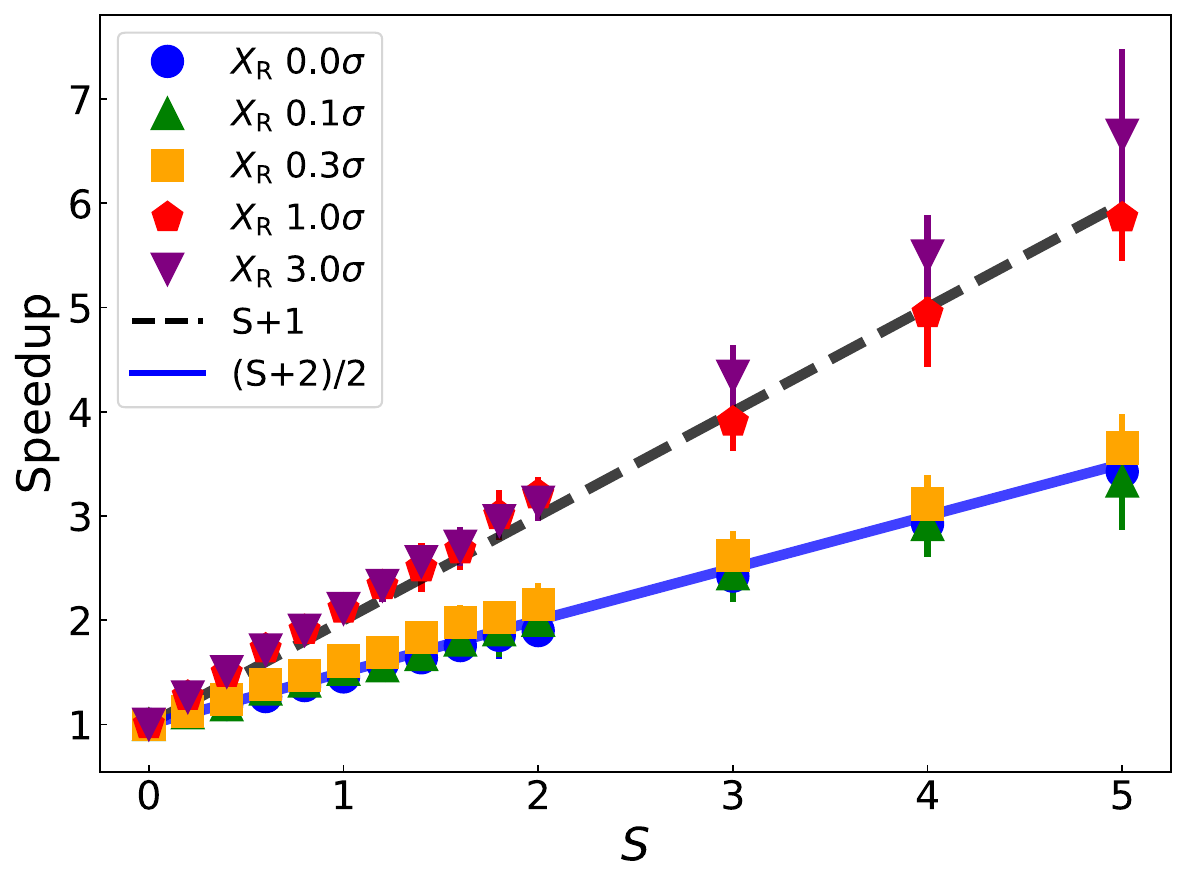}
    \caption{The speedup is presented as a function of $S$ for various initial positions. 
    Dashed lines (dashed black and solid blue) represent the ratio of the first and second moments' timescales, respectively.
    The speedup follows either the first or second moment ratios, depending on initial conditions, with some transition between these two regimes.}
    \label{fig:Xr-Dynamics}
\end{figure}

\textit{Dependence on initial conditions} ---  
So far, we have considered the case where the resetting position $X_r$ coincides with the minimum of the harmonic potential, $X_r=x_0=0$. 
We now turn to a more general setting in which the resetting and initial position are displaced from the minimum ($X_r = x_0 \neq 0$).
This introduces a difference from thermal systems: in equilibrium, the final distribution is independent of the initial condition, whereas in systems under SR, the NESS explicitly depends on the resetting position (see Fig.~\ref{fig:Xr-NESS}).
Interestingly, even when $X_r\neq0$, SR accelerates the relaxation compared to purely thermal dynamics, as shown in Fig.~\ref{fig:Xr-Dynamics}.
Furthermore, two distinct regimes emerge.
For small $X_r$, the speedup in relaxation grows with $S$ as $(S+2)/2$, as observed for $x_0 = 0$, whereas for large $X_r$ it scales as $S+1$. 
This transition can be understood by noting that, for small $X_r$, the first moment of the distribution remains close to zero and therefore the relaxation is governed primarily by the second moment. 
In contrast, for large $X_r$, the first moment becomes significant and, since it relaxes more slowly, dominates the relaxation process (see SI section~\ref{app:moments} for details).
Consequently, applying SR for large $X_r$ results in greater acceleration than for small $X_r$ at the same resetting rate.
\textit{Conclusions} --- 
We have shown that stochastic resetting enhances relaxation towards a NESS for Brownian motion in a harmonic potential; however, it leads to a different steady-state.
The speedup is governed by a single dimensionless parameter $S$ and there is a tradeoff between the extent of spatial exploration and the speedup obtained.
In the control of mechanical systems, this combination of accelerated relaxation and smaller variance may lead to smoother control protocols, with possible applications in soft robotics \cite{haggerty2023control} and active solid actuation \cite{hernandez2024model}.
These insights could inform optimal control strategies~\cite{goerlich_resetting_2024, goerlich2025timeenergytradeoffstochasticresetting, DeBruyneResettingOptimalControl}, search processes~\cite{bressloff_search_2020, singh_backbone_2021, tal-friedman_experimental_2020, pal_random_2024}, and trap-assisted experimental setups.
We focused on a harmonic potential, but the results are more broadly applicable, since any potential can be approximated by a harmonic trap near local minima. 
Future work will extend our analysis to more complex systems with different potentials, such as anharmonic, non-linear, or disordered systems. 
We expect more significant speedups for these scenarios, which have a broader available parameter space, such as in computer simulations~\cite{Church2025Accelerating,blumer_combining_2024,blumer_stochastic_2022}.
We will also explore different resetting protocols, such as space-dependent resetting, or different types of dynamics, such as underdamped or active.
% \remi{Such a tradeoff can have contrasting effects. 
% For example, in sampling methods \cite{blumer_combining_2024}, the reduced variance can bias estimates by under-sampling rare states. 
% Similarly, in the context of population genetics \cite{da2018interplay} the accelerated evolution will come at the risk of eroding diversity.
% In contrast, }
%While our results demonstrated the ability of SR to accelerate relaxation, a general understanding of resetting-induced relaxation remains open.
\section*{Acknowledgments}
Y.R. acknowledges support from the Israel Science Foundation (grants No. 385/21). Y.R., R.G., and N.S. acknowledge support from the European Research Council (ERC) under the European Union’s Horizon 2020 research and innovation programme (Grant agreement No. 101002392). R.G. acknowledges support from the Mark Ratner Institute for Single Molecule Chemistry at Tel Aviv University. 
B.H. acknowledges support by the Israel Science Foundation (grants No.~1037/22 and 1312/22) and the Pazy Foundation of the IAEC-UPBC (grant No.~415-2023). 
 
\section{SI: Steady State under resetting}
\label{app:SR steady-state}
The steady-state distribution for an O-U process characterized by a Diffusion coefficient $D$, frequency $\omega_0$, and undergoing resetting at a rate $r$ is known analytically~\cite{goerlich_taming_2024, pal_diffusion_2015} and is given by: 
\begin{equation}
    P(x)=\sqrt{\frac{a}{\pi p^2}}\Gamma\left(\frac{1}{p}\right)e^{-\frac{ax^2}{2}}\left(ax^2\right)^{-\frac{1}{4}}W_{\frac{1}{4}-\frac{1}{p},\frac{1}{4}}\left(ax^2\right)
\end{equation}\\
where $W_{k,m}(x)$ is the Whittaker W-function, $a=\omega_0/2D$, $b=2\omega_0$, $p=b/r$.

\section{SI: Determining relaxation threshold}
\label{app:threshold}
We define the full relaxation time of a process as the time at which the Kullback–Leibler divergence, $\mathcal{D}_{\rm KL}$, crosses a specified threshold (see Fig.~1 in the main text). This definition enables us to treat, on an equal footing, processes with different functional forms of relaxation and those exhibiting more than one characteristic relaxation time.
As can be expected, the steady-state value of the  $\mathcal{D}_{\rm KL}$ depends on the potential stiffness $\kappa$ and resetting rate $r$. 
Because of the variation in $\mathcal{D}_{\rm KL}$, the threshold is determined adaptively relative to its value at steady-state.
To do so, we quantify the mean and standard deviation of the $\mathcal{D}_{\rm KL}$ at full relaxation.
We set the threshold at 10 standard deviations from the mean value, as this value produced consistent, reproducible results across all systems.

\section{SI: Derivation of the time-dependent moments}
\label{app:moments}

We derive the moments of the probability distribution for a centered $(\langle x\rangle=0)$ Ornstein-Uhlenbeck process under Poisson stochastic resetting with rate $r$, and an arbitrary resetting point $X_r$. 
The renewal equation states that,
\begin{equation}
    \begin{split}
        &P(x,t|x_0) = G_0(x,t|x_0)e^{-rt}+r\int_0^t{G_0(x,t|X_r)e^{-r\tau}d\tau},
    \end{split}
\end{equation}
where $G_0(x,t|x_0)$ is the centered O-U propagator with an initial condition $G_0(x,0) = \delta(x-x_0)$ and , here we enforce $X_r=x_0$ i.e the resetting point coincides with the initial position.
The complete propagator is given in Eq.(2) in the main text.
Our aim is not to solve the last renewal equation for the general time-dependent probability distribution under resetting. 
We will instead calculate the time-dependent first and second moments.
To do so, we multiply both sides by $x^n$ and integrate with respect to $x$. 
This gives us a time-dependent expression for the n-th moment of the distribution under resetting, denoted by $\langle x^n \rangle^{\rm SR}(t)$.

L.H.S:
\begin{equation}
    \begin{split}
        &\int_{-\infty}^\infty x^nP(x,t|x_0)dx = \langle x^n \rangle^{\rm SR}(t)
    \end{split}
\end{equation}

R.H.S:
\begin{equation}
    \begin{split}
        &\left[\int_{-\infty}^\infty x^nG_0(x,t|x_0)dx\right]e^{-rt}\\ 
        &+ r\int_0^te^{-r\tau}\int_{-\infty}^\infty x^n{G_0(x,t|X_r)dx}d\tau\\
        &= \langle x^n \rangle^{\rm eq}(t)e^{-rt}+r\int_0^te^{-r\tau}\langle x^n \rangle^{\rm eq}(\tau)d\tau
    \end{split}
\end{equation}
where $\langle x^n \rangle^{\rm eq}(t)$ is the time-dependent moment of the corresponding thermal system ($r=0$).
Equating the two terms, we arrive at an equation that allows us to compute a general time-dependent moment
\begin{equation}
    \begin{split}
        &\langle x^n \rangle^{\rm SR}(t)=\langle x^n \rangle^{\rm eq}(t)e^{-rt}+r\int_0^te^{-r\tau}\langle x^n \rangle^{\rm eq}(\tau)d\tau.
    \end{split}
    \label{eq:moments-renewal}
\end{equation}
Using Eq.~\ref{eq:moments-renewal} we derive the following expressions for the moments, $\omega_0=\kappa/\gamma$:
\begin{equation}
    \begin{split}
        \langle x \rangle^{\rm SR}(t) &= x_0e^{-(r+\omega_0)t}+\frac{x_0r}{r+\omega_0}(1-e^{-(r+\omega_0)t})\\
        \langle x^2 \rangle^{\rm SR}(t) &= \frac{2D+rx_0^2}{r+2\omega_0}+\frac{2\omega_0 x_0^2-2D}{r+2\omega_0}e^{-(r+2\omega_0)t}
    \end{split}
\end{equation}
In terms of the dimensionless parameters presented in the main text, $S, \tilde x, \tilde t$ and $\tilde x_0$, which is the dimensionless starting position $x_0/\alpha$, we obtain:
\begin{equation}
    \begin{split}
        \langle \tilde x \rangle^{\rm SR}(\tilde t) &= \tilde x_0e^{-(S+1)\tilde t}+\frac{\tilde x_0S}{S+1}(1-e^{-(S+1)\tilde t})\\
        \langle \tilde x^2 \rangle^{\rm SR}(\tilde t) &=\frac{2\tilde x_0^2+2S}{S+2}+\frac{2\tilde x_0^2 - 2S}{S+2}e^{-(S+2)\tilde t}\\
    \end{split}
    \label{eq:time dependent moments SR}
\end{equation}
The timescales of the first and second moments are $(S+1)^{-1}$ and $(S+2)^{-1}$ respectively. 
Whereas for the thermal system, in this dimensionless form the timescales for the moments are $1$ and $1/2$ respectively. 
Thus, the ratios of the moments timescales are $(S+1)/1$ for the mean and $(S+2)/2$ for the second moment.

%\bibliography{export-data}

\end{document}